\documentclass[superscriptaddress,secnumarabic,
amssymb,amsmath,nobibnotes,aps,prd,showkeys,showpacs,nofootinbib]{revtex4-1}%
\usepackage{graphicx}
\usepackage{epsf}
\usepackage{bm}
\usepackage{amsmath}
\usepackage{amsfonts}
\usepackage{amssymb}
\usepackage{epstopdf}
\usepackage{subfigure}
\usepackage{natbib}
\usepackage{color}%
\usepackage{hyperref}
\hypersetup{dvips,dvipdfm,linktoc=page,colorlinks=true,linkcolor=blue,citecolor=red,
filecolor=magenta,urlcolor=blue,bookmarks=true}
\providecommand{\U}[1]{\protect\rule{.1in}{.1in}}
\newcommand{\be}{\begin{equation}}
\newcommand{\ee}{\end{equation}}

\newcommand{\mincir}{\raise
-3.truept\hbox{\rlap{\hbox{$\sim$}}\raise4.truept\hbox{$<$}\ }}
\newcommand{\magcir}{\raise
-3.truept\hbox{\rlap{\hbox{$\sim$}}\raise4.truept\hbox{$>$}\ }}

\begin{document}
\title{Dynamical system analysis of logotropic dark fluid with a power law in the rest-mass energy density}

\author{Goutam Mandal\footnote{Electronic Address: \texttt{\color{blue} gmandal243@gmail.com}}}
\affiliation{Department of Mathematics, University of North Bengal, Raja Rammohunpur, Darjeeling 734013, West Bengal, India}

\author{Sujay Kr. Biswas\footnote{Electronic Address: \texttt{\color{blue} sujaymathju@gmail.com; sujay.math@nbu.ac.in}}}
\affiliation{Department of Mathematics, University of North Bengal, Raja Rammohunpur, Darjeeling 734013, West Bengal, India}

\author{Subhajit Saha\footnote {Electronic Address: \texttt{\color{blue} subhajit1729@gmail.com}}}
\affiliation{Department of Mathematics, Panihati Mahavidyalaya, Kolkata 700110, West Bengal, India}

\author{Abdulla Al Mamon\footnote {Electronic Address: \texttt{\color{blue} abdulla.physics@gmail.com}}}
\affiliation{Department of Physics, Vivekananda Satavarshiki Mahavidyalaya (affiliated
to the Vidyasagar University), Manikpara-721513, West Bengal, India.}


\begin{abstract}


We consider a spatially flat FLRW universe. We assume that it is filled with dark energy in the form of logotropic dark fluid coupled with dark matter in the form of a perfect fluid having a barotropic equation of state. We employ dynamical system tools to obtain a complete qualitative idea of the evolution of such a universe. It is interesting to note that we ought to consider an approximation for the pressure of the logotropic dark fluid in the form of an infinite series so as to be able to construct the autonomous system required for a dynamical system study. This series form provides us with a power law in the rest-mass energy density of the logotropic dark fluid. We compute the critical points of the autonomous system and analyze these critical points by applying linear stability theory. Our analysis reveal a scenario of late-time accelerated universe dominated by the logotropic fluid which behaves as cosmological constant, preceded by an intermediate phase of the Universe dominated by logotropic fluid which behaves as dark matter in the form of perfect fluid. Moreover, it also crosses the phantom divide line.
\end{abstract}

\maketitle

\section{Introduction}

The discovery of cosmic acceleration in 1998 by two independent groups of scientists signifies the birth of a new dawn in Cosmology. However, this discovery brought about a host of new challenges for the cosmological community since the contemporary Standard Model of Cosmology suggested a decelerating universe. This unexpected discovery led most cosmologists to look for modifications into the contemporary theory. The first attempt saw the revival of the cosmological constant\footnote{Interested readers may see the paper by Chavanis \cite{Chavanis1} for some interesting anecdotes on the cosmological constant.} $\Lambda$, which was once introduced by Einstein in his field equations in order to obtain an infinite homogeneous static universe but was later forced to withdraw it when Hubble discovered in 1929 that our Universe is rather an expanding one. It is quite evident that a positive cosmological constant can take into account the presently observed accelerating universe. Interested readers may look into the extensive review on the cosmological constant by Padmanabhan \cite{Padmanabhan1}. The present Standard Model of Cosmology, also known as the Lambda cold dark matter ($\Lambda$CDM) model, consists of a presureless (cold) dark matter (DM) and a positive cosmological constant ($\Lambda >0$). $\Lambda$ is one of a class of many such fluids which have been collectively termed as dark energy (DE). The term ``dark" refers to the fact that the nature and origin of this exotic class of fluids is still elusive to Cosmologists even after more than two decades of solid research both theoretically as well as observationally. But why do we need other forms of dark energy if $\Lambda$ describes the Universe so well? To answer this question, we need to look at the physical interpretation of $\Lambda$. The cosmological constant is interpreted in terms of the energy density of the vacuum. This leads to a discrepancy of $10^{123}$ orders of magnitude in the value of $\Lambda$ since quantum field theory predicts that the energy density of the vacuum must be of the order of the Planck density. This discrepancy is known as the cosmological constant problem in Cosmology. To alleviate this problem, a group of authors \cite{Caldwell1,Starobinsky1,Saini1} abandoned $\Lambda$ altogether and instead introduced a dark energy fluid having a time-varying density associated to a scalar field known as quintessence. Later, Kamenshchik {\it et al.} \cite{Kamenshchik1} proposed a unified model of DM and DE by considering a fluid known as the Chaplygin gas. This was the first instance of assuming a single dark fluid to reproduce the entire evolutionary history of the Universe and also to explain the present accelerating phase. Nevertheless, the Chaplygin gas model does not fit the observations well \cite{Sandvik1,Zhu1} which is why several generalizations of the Chaplygin gas have been considered in the literature \cite{Bilic1,Fabris1,Bento1,Benaoum1,Gorini1,Bento2,Debnath1,Zhang1,Saha1}. Among them, the modified Chaplygin gas (MCG) and the generalized Chaplygin gas (GCG) have been widely discussed in the literature although they have also failed to be accepted as promising candidates when observations were taken into account \cite{Benaoum1}. Chavanis \cite{Chavanis2,Chavanis3,Chavanis4} considered a heuristic unification of DM and DE by assuming an equation of state associated with a constant negative pressure. This is a simple model which is consistent with observations but it has several drawbacks since it has no free parameter and it fails to describe DM halos. In this regard, Chavanis \cite{Chavanis1} argues that in order to realize an effective unification of DM and DE, one must consider a fluid with an equation of state which increases slowly with the energy density. This observation leads naturally to a polytropic equation of
state where the pressure varies with the density in a power-law fashion corresponding to a small index $\gamma \simeq 0$. Tooper \cite{Tooper1,Tooper2} was the first to realize the possibility of two types of polytropic equations of state in the context of general relativistic stars. Their cosmological implications have been studied by Chavanis \cite{Chavanis2,Chavanis3,Chavanis4}. It is worthwhile to mention here that a polytropic equation of state with a $\gamma \neq 0$ can be obtained from either the Gross-Pitaevskii equation (GPE) or the Klein-Gordon equation (KGE). However, there is no physical justification behind a non-zero index. As expected, the trivial case $\gamma = 0$ corresponds to the $\Lambda$CDM model but, in that case, the GPE and the KGE degenerate. With this view in mind that the existing unified models of DM and DE are quite unsuccessful in the context of Cosmology, Chavanis \cite{Chavanis1} investigated a new unification of DM and DE where the pressure varies as a logarithm of the rest-mass energy density. This equation of state (EoS), commonly known as the logotropic equation of state, was first conceived by McLaughlin and Pudritz \cite{McLaughlin1} in astrophysics to understand the internal structure and the average properties of molecular clouds and clumps. The cosmological aspects of this EoS has already been studied in Refs. \cite{Chavanis1,Chavanis10,Chavanis11,Mamon1}. Using the
latest Pantheon SNIa and cosmic chronometer datasets, very recently, Mamon and Saha \cite{Mamon1} have shown that the logotropic model may alleviate the $H_{0}$ tension problem. The logotropic model has several nice properties \cite{Chavanis1} and we enumerate them below---
\begin{enumerate}
\item [a.] The universal rotation curve generated by logotropic DM halos matches the observational Burkert profile \cite{Burkert1} up to the halo radius.
\item [b.] If all the DM halos are assumed to have the same logotropic temperature $A$ (see Section \ref{sec2}), then their surface density is found to be a constant, a result which is in excellent agreement with observations \cite{Kormendy1,Spano1,Donato1}.
\item [c.] Gravitational thermodynamics predicts a value of the dimensionless logotropic temperature $B$ ($B=A/\rho_{\Lambda}c^2$) to be within $0.404$ \cite{Mamon1}. Chavanis \cite{Chavanis1} has obtained the exact value $B=3.53 \times 10^{-3}$ from galactic observations. This shows that the logotropic equation of state is consistent with thermodynamics. 
\end{enumerate}
These interesting properties reveal that the logotropic equation of state deserves a deeper study. In this paper, we wish to investigate the implications of the logotropic dark fluid coupled with DM in the context of dynamical system analysis. But, there is a twist here. We ought to consider an approximation for the pressure of the logotropic dark fluid in the form of an infinite series so as to be able to construct the autonomous system required for a dynamical system study. This series form will provide us with a power law in the rest-mass energy density of the logotropic dark fluid. It may be so that the failure to anticipate this trick has not prompted researchers to undertake a dynamical system analysis of this model till now.
Usually, when the cosmological evolution equations are complicated in nature and are unable to provide the exact analytical solutions, then the dynamical system tools have to be employed to overcome the complexity and to achieve the solutions qualitatively. This method has become an attractive and promising one in recent years since it gives us the complete evolutionary scheme of the universe through analysis of the critical points of the autonomous system. As a result, dynamical systems analysis has been extensively performed in various cosmological models like scalar field dark energy models, interacting dark energy models, modified gravity theory models etc. see for instance in Refs. \cite{M.Khurshudyan2015,S.Kr.Biswas2015a,S.Kr.Biswas2015b,C.G.Bohmer2008,N.Tamanini2015,Xi-ming Chen2009,Copeland1,Fang1, Odintsov2017,Oikonomou2018,Odintsov2018a,Kleidis2018,Odintsov2018b,Aljaf, S.K.Biswas2021} where various cosmological models have been studied in dynamical systems perspectives. Recently, dynamical systems analysis in various modified gravity theories and dark energy models have been studied exhaustively in a review article in the Ref. \cite{Bahamonde2018}.\\

In this paper, the autonomous system is perturbed upto the first order around the critical points and the Hartman-Grobman theorem is applied to find the nature of the critical points. Then, the stability of the points is obtained by finding eigenvalues of the perturbed matrix at the critical points. Our study reveals cosmologically viable scenarios in the context of dynamical analysis such as critical points representing late-time dark energy dominated accelerated evolution of the universe preceded by a matter-dominated intermediate phase of the universe, where logotropic fluid mimics dark energy as well as dark matter.\\ 

The paper is organized as follows. Section \ref{sec2} is concerned with the basic setup of the logotropic model in Cosmology. Section \ref{sec3} and its subsections deal with the dynamical system analysis of the model. A summary of the work and a discussion of the results obtained can be found in Section \ref{sec4}.
\section{Basic equations in the Logotropic Model}
\label{sec2}
We assume a spatially flat, homogeneous and isotropic FLRW universe filled with a perfect fluid having energy density $\epsilon(t)$, rest mass density $\rho(t)$ and isotropic pressure $p(t)$. Then, the Friedmann equation and the
energy conservation equation can be obtained as \cite{refwbbook}
\begin{equation}\label{eqam1}
H^{2}=\left(\frac{\dot{a}}{a}\right)^{2}=\frac{8\pi G}{3}\epsilon,
\end{equation}
\begin{equation}\label{eqam2}
\frac{d\epsilon}{dt}+3H\left(\epsilon + \frac{p}{c^{2}}\right)=0,
\end{equation}
respectively. In the above expressions, $H=\frac{\dot{a}}{a}$ denotes the Hubble parameter, $a(t)$ denotes the scale factor of the universe and an overhead dot represents derivative with respect to the cosmic time $t$. Also, the symbols $G$ and $c$ represent the universal gravitational constant and the velocity of light, respectively. According to thermodynamical laws in logotropic Cosmology, one can obtain the energy density of logotropic DE as the combination of rest mass energy density and the internal energy density \cite{Chavanis1,Chavanis10,refwbbook}
\begin{equation}\label{total-energy}
	\epsilon=\rho c^{2}+u(\rho),
\end{equation}
where $u(\rho)=\rho \int^{\rho} \frac{p(\rho')}{\rho'^{2}} d\rho'=-p(\rho)-A$, is the internal energy density and $A$ is the logotropic temperature. It is evident from Eq. (\ref{total-energy}) that the energy density is the sum of the rest-mass energy $\rho c^{2}$ and the internal energy $u(\rho)$. Note that $\rho c^{2}$ is always positive while $u(\rho)$ may be positive or negative, and $\epsilon$, the logotropic energy density is always positive in its evolution. Also, the pressure of the logotropic single dark fluid can be evaluated as \cite{Chavanis1,Chavanis10}
\begin{equation} \label{pressure}
p=A \ln \left(\frac{\rho}{\rho}_{*}\right),~~~~~ A\ge 0,
\end{equation}
with $\rho_{*}=5.16\times10^{99} ~~gm/m^{3}$ is the Plank density. Eq. (\ref{pressure}) is known as the logotropic EoS and the fluid which obeys this EoS will be called the logotropic dark fluid. As shown in \cite{Chavanis1}, the
rest-mass energy of the logotropic dark fluid mimics DM and its internal energy mimics DE. Therefore, Eq. (\ref{pressure}) leads to a natural and physical unification of DM and DE, and interprets their mysterious nature.\\

Now, without any loss of generality, we assume that the Planck density and the velocity of light to be unity (i.e., $\rho_{*}=1$ and $c=1$). Keeping in line with our earlier discussion, we now obtain a power-law form for the pressure $p$ in terms of the rest-mass energy density $\rho$ which is given by
 \begin{equation}\label{logotropic-pressure}
 	p=-\frac{A}{2} \rho^{2}+2A\rho-\frac{3A}{2}.
 \end{equation}
For simplicity, we have kept terms up to the second degree in $\rho$. The logotropic fluid energy density, $\epsilon$, now takes the form
\begin{eqnarray}
	\epsilon &=& \rho-p-A \nonumber \\
			 &=& \frac{A}{2}+\rho(1-2A)+\frac{A}{2}\rho^{2}.
\end{eqnarray}
The energy conservation equation for logotropic fluid is given by
\begin{equation}\label{energy-conservation-logotropic-fluid}
	\dot{\epsilon}+3H(\epsilon+p)=0
\end{equation}
where $p$ is the logotropic pressure defined by Eq. (\ref{logotropic-pressure}).
Now considering the Universe is filled with logotropic fluid as DE candidate along with DM in the form of perfect fluid satisfying barotropic equation of state $\omega_{m}=\frac{p_m}{\rho_{m}}$, where perfect fluid describes dust for $\omega_{m}=0$, radiation for $\omega_{m}=\frac{1}{3}$ and stiff matter for $\omega_{m}=1$, the modified Friedmann and the acceleration equations become
\begin{equation}\label{Friedmann-equation}
H^{2}=\frac{1}{3}(\epsilon+\rho_{m})
\end{equation}
and 
\begin{equation}\label{acceleration-equation}
\dot{H}=-\frac{1}{2}(\rho_{m}+\epsilon+\omega_{m}\rho_{m}+p)
\end{equation}
respectively. Also, the conservation equation for DM is given by
\begin{equation}\label{energy-conservation}
	\dot{\rho}_{m}+3H\rho_{m}(1+\omega_{m})=0,
\end{equation}
where the EoS parameter for DM ranges as  $0\leq \omega_{m}\leq 1.$
The density parameter for the logotropic fluid and that for DM can be defined as $\Omega_{\epsilon}=\frac{\epsilon}{3H^2}$ and $\Omega_{m}=\frac{\rho_{m}}{3H^2}$ respectively, thus leading to the Friedmann constraint as $1=\Omega_{\epsilon}+\Omega_{m}.$
The argument behind considering the DM candidate as an individual matter content of the universe is that the DM has a different evolutionary scheme and does not affect the evolution of logotropic fluid as they have separate conservation equations. Apart from this, a complete qualitative behavior (i.e., from early DM dominated evolution of universe to DE dominated late phase of universe) of the system can be achieved where the DM plays an important role in the perspective of dynamical system analysis.



\section{Dynamical System Analysis of the Logotropic Model}
\label{sec3}

To get an idea about the qualitative behavior of evolution, we introduce new dynamical variables as:
\begin{equation}
x=\frac{\rho}{3H^{2}},~~~~ y=\frac{\rho_{m}}{3H^{2}}.
\end{equation}
Note that the energy density $\rho$ of rest mass and energy density $\rho_{m}$ of DM are not interrelated because none of them is present explicitly in the expression of each and as a result the dynamical variables $x$ and $y$ are independent. Therefore, a valid autonomous system of ODEs of $x$ and $y$ can be constructed. The acceleration equation (\ref{acceleration-equation}) (with Friedmann equation (\ref{Friedmann-equation})) can be written in terms of $x$ and $y$ as
\begin{equation}
	\frac{\dot{H}}{H^{2}}=-\frac{1}{2}\left(3+3\omega_{m}y-\frac{9A}{2}x^{2}H^{2}+6Ax-\frac{3A}{2H^{2}}\right).
\end{equation}
System of ODEs of the variables $x$ and $y$ is:
\begin{eqnarray}
\begin{split}
   \frac{dx}{dN}& = 3 \left[ x \left\lbrace 3 A H^2 x^2+(3-4 A) x+y \left(\omega _m+3\right)-2\right\rbrace -\frac{3 A H^2 x^2+(3-4 A) x+2 (y-1)}{3 A H^2 x-2 A+1}\right] ,& \\
   \frac{dy}{dN}& =3 y \left\lbrace 3 A H^2 x^2+(3-4 A) x+(y-1) \left(\omega _m+3\right)\right\rbrace, &~~\label{autonomous1}
\end{split}
\end{eqnarray}
where the e-folding parameter $ N=\ln a $ is taken as independent variable for the above system.
Now, in presence of the Hubble parameter $H(t)$ explicitly in the expressions of right hand sides of both of equations, the system of ODEs (\ref{autonomous1}) in ($x-y$) does not lead to an autonomous system. That is here the Hubble parameter $H(t)$ is considered to be present as a cosmological variable but not a dynamical variable. We choose the Hubble parameter as a dynamical variable $H(N)$. This type of consideration employed in understanding cosmological dynamics using dynamical systems have been made several times in the literature, see for example in the Ref. \cite{Oikonomou2019}. Then, with an additional equation of $\frac{dH}{dN}$, the system (\ref{autonomous1}) will be in closed form and leading to an autonomous system of $x(N)$, $y(N)$, and $H(N)$ in the following:
  \begin{eqnarray}
  	\begin{split}
  		\frac{dx}{dN}& = 3 \left[ x \left\lbrace 3 A H^2 x^2+(3-4 A) x+y \left(\omega _m+3\right)-2\right\rbrace -\frac{3 A H^2 x^2+(3-4 A) x+2 (y-1)}{3 A H^2 x-2 A+1}\right] ,& \\
  		\frac{dy}{dN}& =3 y \left\lbrace 3 A H^2 x^2+(3-4 A) x+(y-1) \left(\omega _m+3\right)\right\rbrace  ,& \\
  		\frac{dH}{dN} & =-\frac{3H}{2} \left\lbrace 3 A H^2 x^2+(3-4 A) x+y \left(\omega _m+3\right)-2\right\rbrace .&~~\label{autonomous}
  	\end{split}
  \end{eqnarray}
The Friedmann equation (\ref{Friedmann-equation}) gives the constraint in terms of $x$, $y$, $H$ in the phase space as:
\begin{equation}
	1=\frac{A}{6H^{2}}+(1-2A)x+\frac{3}{2}Ax^{2}H^{2}+y,
\end{equation} 
Now, using the dynamical variables the cosmological parameters can be expressed in the following forms.\\\\
The density parameter for DM:
\begin{equation}
\Omega_{m}=\frac{\rho_{m}}{3H^2}=y
\end{equation}
Energy conditions in cosmology implies the restrictions in phase space variables i.e., 
$0\leq\Omega_{m}\leq1 \Longrightarrow 0\leq y\leq1.$
The density parameter for logotropic fluid (since from Friedmann equation (\ref{Friedmann-equation}) we have the constraint: $1=\Omega_{\epsilon}+\Omega_{m}$):
\begin{equation}
\Omega_{\epsilon}=\frac{\epsilon}{3H^2}=\frac{A}{6H^{2}}+(1-2A)x+\frac{3}{2}Ax^{2}H^{2}=1-y
\end{equation}
It should be noted that since the logotropic pressure ($p$) as well as logotropic energy density ($\epsilon$) are considered to be explicit functions of rest mass density $\rho$, the dynamical system shows the same nature for the variable $x$ which is expressed in terms of $\rho$ normalized over Hubble scale. Therefore, $x$ will appear naturally in the physical parameters.\\\\
The equation of state parameter for logotropic fluid:

\begin{equation}
\omega_{\epsilon}=\frac{p}{\epsilon}=\frac{3 A H^2 x^2+(3-4 A) x+3 y-3}{1-y}
\end{equation}
The effective equation of state parameter:
\begin{equation}
\omega_{eff}=\frac{p+p_m}{\epsilon+\rho_{m}}=3 A H^2 x^2+(3-4 A) x+y \left(\omega _m+3\right)-3
\end{equation}
The deceleration parameter takes the form:
\begin{equation}
q=-1-\frac{\dot{H}}{H^2}=\frac{3}{2} \left\lbrace 3 A H^2 x^2+(3-4 A) x+y \left(\omega _m+3\right)-2\right\rbrace -1.
\end{equation}
 There will be acceleration for $q<0$ or $\omega_{eff}<-\frac{1}{3}$.

\subsection{Critical points and phase space analysis} 

In this section, we shall discuss the phase space analysis of critical points obtained from the autonomous system (\ref{autonomous}) by equating the right part of it to zero. The critical points and corresponding cosmological parameters are shown in tabular form in the Table \ref{physical-parameters}. The linear stability theory is applied to examine the nature of critical points by giving a small (up to first order) perturbation around the critical points.  Eigenvalues of the linearized jacobian matrix are shown in the Table \ref{eigen-value}. \\

\begin{figure}
	\centering
	\subfigure[]{%
		\includegraphics[width=5cm,height=6cm]{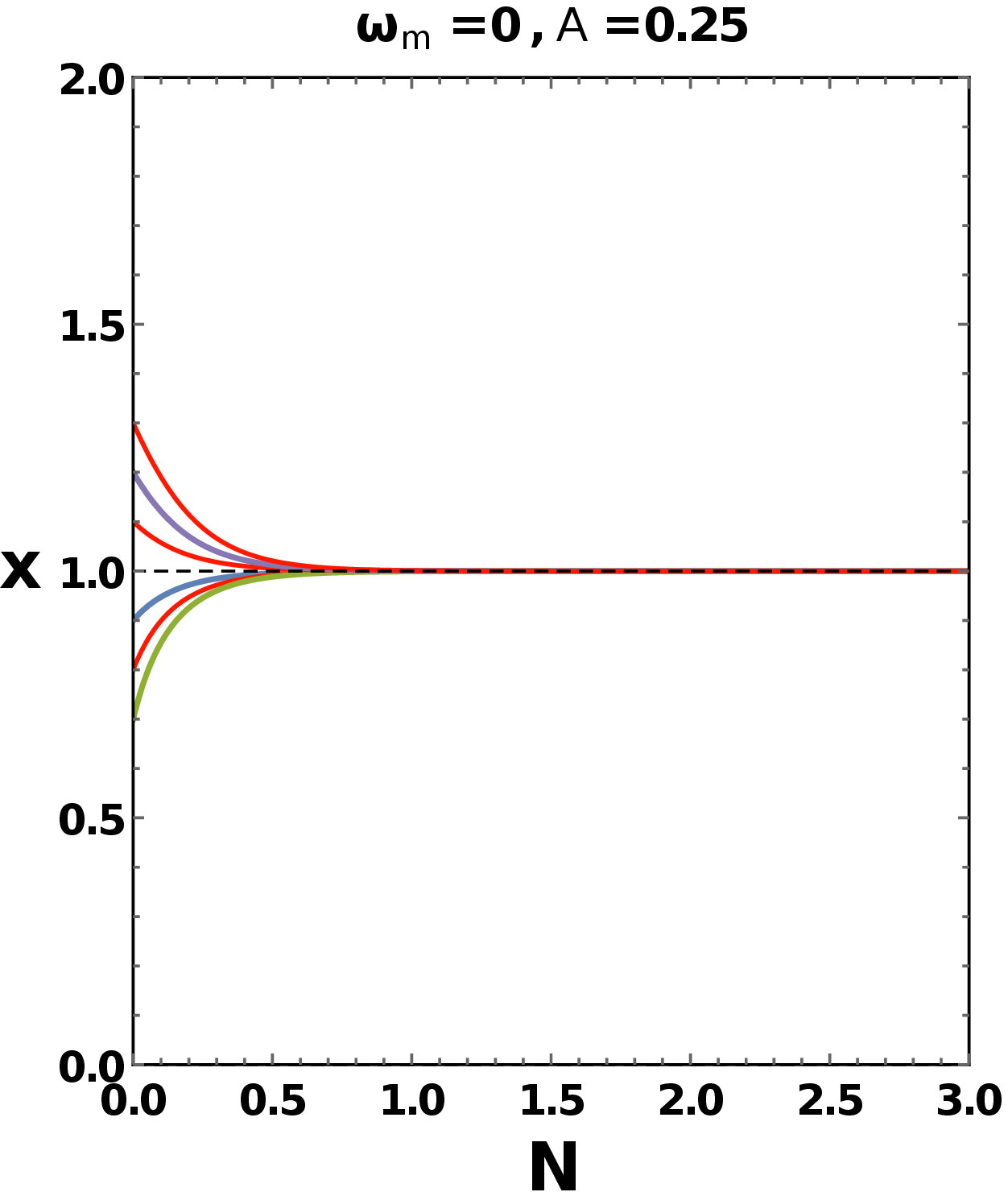}\label{fig:Stable_Nx}}
	\qquad
	\subfigure[]{%
		\includegraphics[width=5cm,height=6cm]{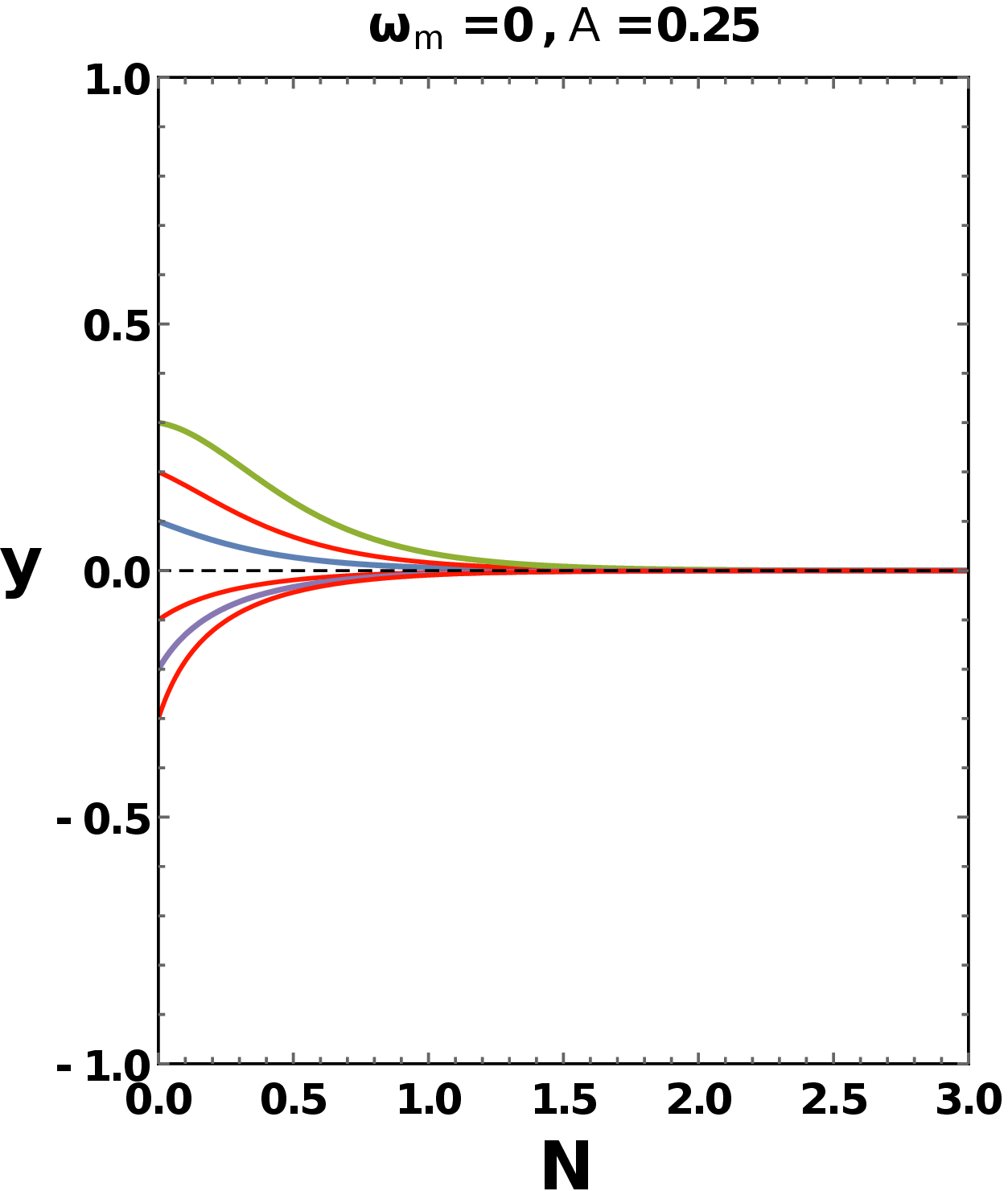}\label{fig:Stable_Ny}}
	\qquad
	\subfigure[]{%
		\includegraphics[width=5cm,height=6cm]{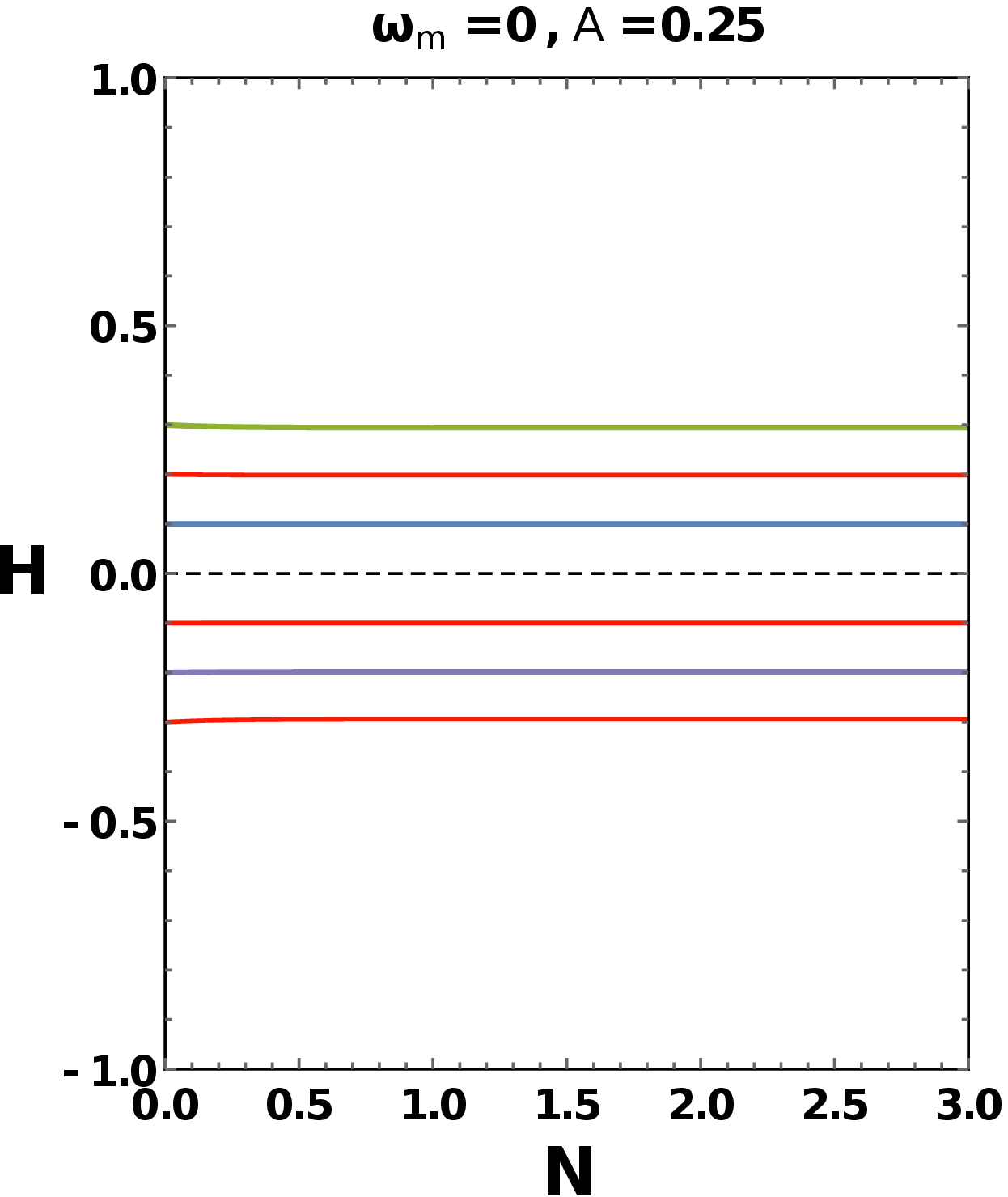}\label{fig:Stable_NH}}
	\caption{The figures show the phase space projection of perturbation of the autonomous system (\ref{autonomous}) along the x,y and H axes for the critical point $C$ for parameter values $\omega_{m}=0$ and $A=0.25$. In panel (a) and (b) perturbations come back but in panel (c) perturbations do not come back. This indicates that the critical point $C$ is unstable for parameters value $\omega_{m}=0$ and $A=0.25 $. }
	\label{C-perturbation}
\end{figure}

\begin{itemize}
\item  I. Critical Points : $ M = (0,1,0) $ always exists for $A\neq \frac{1}{2}$ in the phase space $(x,y,H)$. The point is completely DM dominated ($\Omega_{m}=1$) solution and there is no energy contribution from rest mass energy density towards the logotropic dark fluid (DE), so it is absent here ($\Omega_{\epsilon}=0$). The DE equation of state is undefined for this point. There is always deceleration near the point(since $q>0$). The point is hyperbolic in nature since all the eigenvalues are non-zero (see Table \ref{eigen-value}). The point will be completely dust dominated saddle like decelerated solution when $\omega_{m}\longrightarrow 0$.

\item  II. Critical Point : $ B = (\frac{1}{1-2A},0,0) $ will always exist for $A\neq \frac{1}{2}$ in phase space. The point corresponds to a complete logotropic fluid dominated ($\Omega_{\epsilon}=1$) solution in the phase space where DM is absent ($\Omega_{m}=0$). Therefore, contribution of rest mass energy density is occurring towards the logotropic fluid. There exists acceleration near the point for $A>-\frac{1}{4}$ and phantom crossing ($\omega_{eff}<-1$) behavior is found for the restriction $A>\frac{1}{2}$. The logotropic fluid for this critical point behaves as dust ($\omega_{\epsilon}\approx 0$) for $A\approx 0$  and as a result, there exists an ever decelerating phase near the point. From the Table \ref{eigen-value}, we observe that the point is hyperbolic type and is always unstable saddle like in nature (since one of the eigenvalues is always positive in the phase space). The point shows a solution dominated by the logotropic fluid which mimics as dust giving decelerated expansion of the universe.

\item  III. Critical Point : $ C=(\frac{2}{3-4A},0,0) $ is also a logotropic fluid (DE) dominated solution where energy density is contributed from rest mass energy density only. The point always exists in phase space except for $A=\frac{3}{4}$. DM is absent here. For this critical point, logotropic fluid behaves as cosmological constant like fluid (since $\omega_{\epsilon}=-1$). The point is always accelerating since $q=-1$. The point is non-hyperbolic in nature since one of eigenvalues is zero in phase space (see Table \ref{eigen-value}).
There exist two non-empty stable sub-manifold in the negative eigen directions for parameter restrictions $\omega_{m}>-1$ and $A<\frac{1}{2}$. So it has the nature of de Sitter expansion of the universe ($\Omega_{\epsilon}=1, \Omega_{m}=0, \omega_{eff}=q=-1$). Due to non-hyperbolic in nature, the linear stability theory fails to describe the complete dynamics of the critical points. Generally, the Center Manifold theory can be employed to achieve the stability criteria of the point. However, the stability of this type of point can also be achieved by numerically perturbing the solution trajectories near the point as $N\longrightarrow \infty$ and this method is now a promising one in the literature (see for example \cite{JDUTTA2019}). In figure (\ref{C-perturbation}) the trajectories of the solutions are plotted numerically for $\omega_{m}=0$ and $A=0.25$ along $x$ axis, $y$ axis and $H$ axis where the initial conditions are chosen very close to the coordinates of the point. The figures (\ref{fig:Stable_Nx}) and (\ref{fig:Stable_Ny}) show that the trajectories approach along $x=1$ line and $y=0$ axis as $N\longrightarrow \infty$ where as figure (\ref{fig:Stable_NH}) shows that the solution trajectories starting from a point remains parallel to each other but does not converge along $H=0$. So it can be concluded that the critical point $C$ is not a stable point, but it is a saddle like unstable point and the point behaves as a de Sitter like solution having transient evolution of the universe. 


\item  IV. Set of Critical Points : $ D=(\frac{-3+4A+\sqrt{24 H_c^2 A+16 A^2-24 A+9}}{6 H_c^2 A},0,H_c) $ exists for\\
(1). $A<0~~\mbox{and}~~ \left(\frac{4 A-3}{2 \sqrt{6} \sqrt{-A}}\leq H<0~~\mbox{or}~~ 0<H\leq \frac{3-4 A}{2 \sqrt{6} \sqrt{-A}}\right) ,$ or\\
(2). $A>0~~\mbox{and}~~ H\neq 0$. It is a set of critical points on the plane $y=0$. It is a non-isolated set of points and one of eigenvalues is zero (see Table \ref{eigen-value}). So, the set of points is non-hyperbolic in nature in the phase space. Also, non-hyperbolic critical set with exactly one eigenvalue zero is called normally hyperbolic set \cite{Coley,Sujay2017} and the stability of normally hyperbolic set is determined by the sign of remaining non-zero eigenvalues. The set is completely logotropic fluid dominated ($\Omega_{\epsilon}=1$) because total energy density is contributed from the rest mass energy density and so DM is absent here. Logotropic fluid behaves as cosmological constant ($\omega_{\epsilon}=-1$). Acceleration is always possible near the set of points since $q=-1$ and $\omega_{eff}=-1$. Depending on some parameter restrictions, the set of points will be stable in the phase space for\\
(1). $A<0~~\mbox{and}~~ \left(-\frac{\sqrt{-2 A^2+3 A-1}}{\sqrt{3A}}<H<0~~\mbox{or}~~ 0<H<\frac{\sqrt{-2 A^2+3 A-1}}{\sqrt{3A}}\right) $, or\\
(2). $0<A\leq \frac{1}{2}~~\mbox{and}~~ (H<0~~\mbox{or}~~ H>0) $, or\\
(3). $\frac{1}{2}<A<1~~\mbox{and}~~ \left(H<-\frac{\sqrt{-2 A^2+3 A-1}}{\sqrt{3A}}~~\mbox{or}~~ H>\frac{\sqrt{-2 A^2+3 A-1}}{\sqrt{3A}}\right) $, or\\
(4). $A\geq 1~~\mbox{and}~~ (H<0~~\mbox{or}~~ H>0)$.\\
Therefore, the set of points show the late-time logotropic dominated accelerating stable attractor in the phase space.\\

\item  V.Set of Critical Points : $ E=(\frac{-3+4A-\sqrt{24 H_c^2 A+16 A^2-24 A+9}}{6 H_c^2 A},0,H_c) $  has the similar nature with the set D.
It exists for \\
(1). $A<0~~\mbox{and}~~ \left(\frac{4 A-3}{2 \sqrt{6} \sqrt{-A}}\leq H<0~~\mbox{or}~~ 0<H\leq \frac{3-4 A}{2 \sqrt{6} \sqrt{-A}}\right) ,$ or\\
(2). $A>0~~\mbox{and}~~ H\neq 0$. The set is completely logotropic fluid dominated solution. Acceleration is always possible since $q=-1$ for the set of points. The set of points is non-hyperbolic in nature with exactly one vanishing eigenvalue showing that the set is normally hyperbolic set. Stability of the set is determined by sign of the remaining non-zero eigenvalues. The set is always unstable (saddle) in nature. As a result, it fails to describe the late time solution in the perspective of dynamical system analysis.  Therefore, the normally hyperbolic critical set $E$ is less interested in cosmological view point. \\

\item  VI.  Set of Critical Points : $ F=(x_{c},1-\frac{x_{c}}{1+\omega_{m}},0) $ will exist for $0\leq \omega_{m}\leq1$ and $0\leq x_{c} \leq \omega_{m}+1$ and $\omega_{m}=\frac{2A}{1-2A}$. The set of points is the combination of both logotropic fluid and DM and the ratio of DE and DM is:
$\frac{\Omega_{\epsilon}}{\Omega_{m}}=\frac{x_c}{1+\omega_{m}-x_c}$. In the evolution of this solution (set of critical points)energy density of logotropic fluid is contributed from rest mass density while at the same time the energy density of matter is contributed from individual energy density of DM. Further, When $x_c\longrightarrow 0$, the point will be completely DM dominated and then it will have the similar nature with the point $M$. Logotropic fluid behaves here as a perfect fluid since $\omega_{\epsilon}=\omega_{m}$ and it will be dust for $\omega_{m}=0$, and radiation for $\omega_{m}=\frac{1}{3}$, but can never behaves as DE fluid. The set is non-hyperbolic in nature since one of eigenvalues is zero, and so it is normally hyperbolic set. This is always unstable (saddle like) point in the phase space since one of non-zero eigenvalues is always positive. It cannot represent the late time solution. Therefore, the critical point shows the transient evolution of the universe.  \\
\end{itemize}

\begin{table}[tbp] \centering
\caption{The Critical Points and the corresponding physical parameters for the logotropic model.}%
\begin{tabular}
[c]{ccccccc}\hline\hline
\textbf{Critical Points}&$(\mathbf{x,y,H})$& $\mathbf{\Omega_{\epsilon}}$ &
 $\mathbf{\Omega_{m}}$ & $\mathbf{\omega_{\epsilon}}$ & $\mathbf{{\omega}_{eff}}$ &
\textbf{q}\\\hline
$M$ & $(0,1,0)$ &  $ 0 $ &
$1$ & $\frac{0}{0}$ & $\omega_{m}$ & $\frac{1}{2}(1+3\omega_{m})$\\
$B$ & $(\frac{1}{1-2A},0,0)$ & $ 1 $ & $0$ &
 $ \frac{2A}{1-2A}$ & $ \frac{2A}{1-2A} $ & $ \frac{1+4A}{2-4A}  $ \\
$C$ & $(\frac{2}{3-4A},0,0)$ & $ 1 $ &
 $ 0 $ & $ -1 $ & $ -1 $ & $-1$ \\
$D$ & $(\frac{-3+4A+\sqrt{16 A^2+24 A H_c^2-24 A+9}}{6 A H_c^2},0,H_c)$ & $ 1  $ &
 $ 0 $ & $ -1 $ & $ -1 $ & $ -1 $ \\

$E$ & $(\frac{-3+4A-\sqrt{16 A^2+24 A H_c^2-24 A+9}}{6 A H_c^2},0,H_c)$& $ 1 $ &
 $ 0 $ & $ -1 $ & $ -1 $ & $ -1 $ \\
$F$ & $(x_{c},1-\frac{x_{c}}{1+\omega_{m}},0)$ & $\frac{x_{c}}{1+\omega_{m}}$ & $1-\frac{x_{c}}{1+\omega_{m}}$ & $\omega_{m}$ & $\omega_{m}$ & $\frac{1}{2} \left(3 \omega _m+1\right)$
 \\\hline\hline
\end{tabular}
\label{physical-parameters}
\end{table}%
%
\begin{table}[tbp] \centering
\caption{The eigenvalues of the linearized system  for the logotropic model.}%
\begin{tabular}
[c]{ccccccc}\hline\hline
\textbf{Critical Points} & $\mathbf{\lambda_{1}}$ &
 $\mathbf{\lambda_{2}}$ & $\mathbf{\lambda_{3}}$ \\\hline
$M  $ & $ -\frac{3}{2}(1+\omega_{m}) $ &  $ 3(1+\omega_{m})$ & $ \frac{6 A}{2 A-1}+3 \omega _m $ \\
$B  $ & $ \frac{3}{1-2A} $ &  $ \frac{3}{4A-2} $ & $ \frac{6 A}{1-2A}-3 \omega _m $ \\
$C  $ & $  0  $ &
 $ \frac{3}{2A-1} $ & $-3(1+\omega_{m})$ \\
$D  $ & $ 0  $ & $ \frac{6}{1-\sqrt{8 A \left(3 H_{c}^2+2 A-3\right)+9}}  $ &
 $ -3 \left(\omega _m+1\right) $  \\
 $E  $ & $ 0 $ & $ \frac{6}{1+\sqrt{8 A \left(3 H_{c}^2+2 A-3\right)+9}}  $ &
 $ -3(1+\omega_{m}) $ \\
 $F$ & $0$ & $-\frac{3}{2} \left(\omega _m+1\right)$ & $3 \left(\omega _m+1\right)$ \\

 \\\hline\hline
\end{tabular}
\label{eigen-value}
\end{table}%


\begin{figure}
	\centering
	\subfigure[]{%
		\includegraphics[width=8.5cm,height=8.5cm]{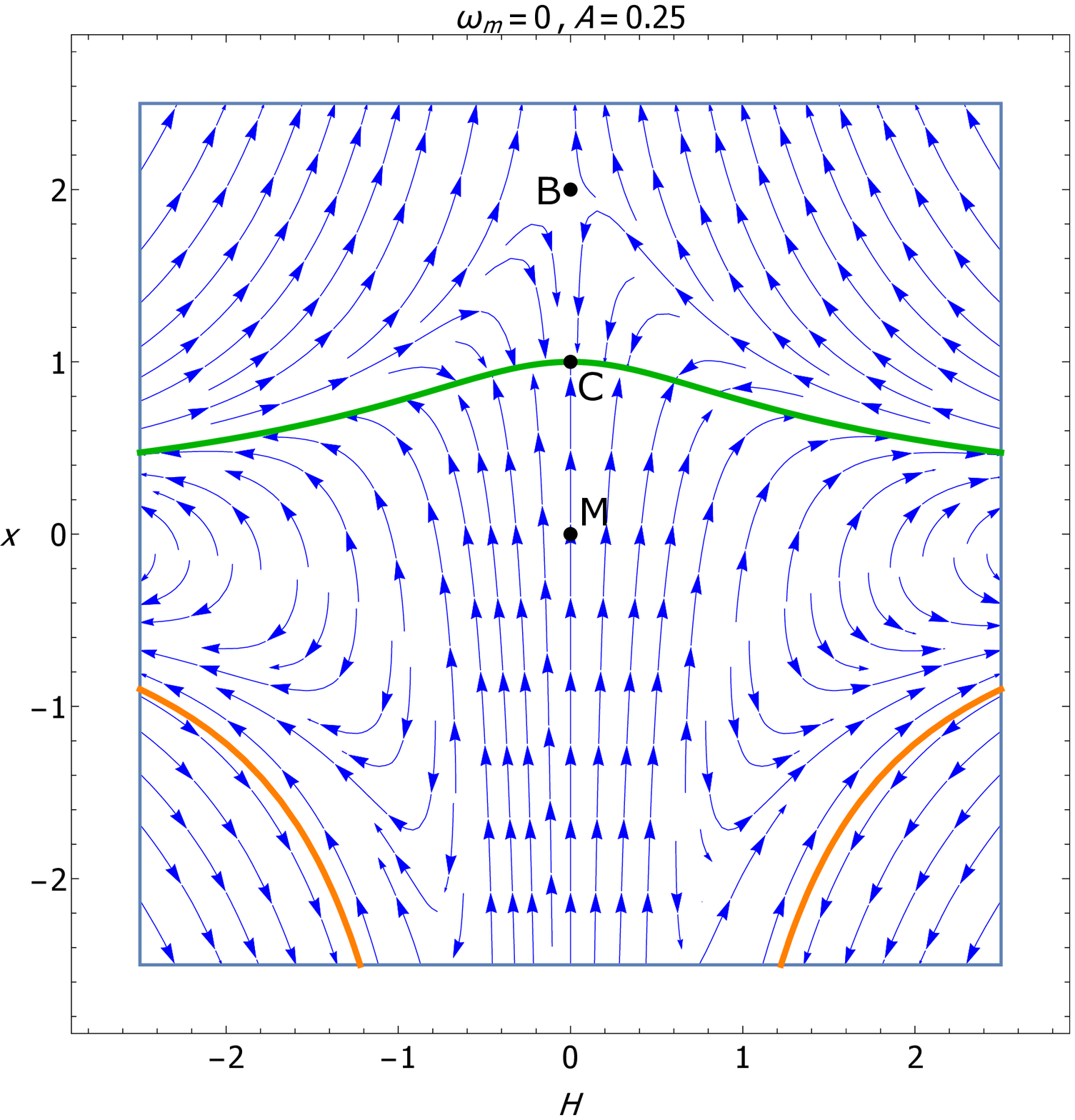}\label{fig:Stable_HX}}
	\qquad
	\subfigure[]{%
		\includegraphics[width=8.5cm,height=8.5cm]{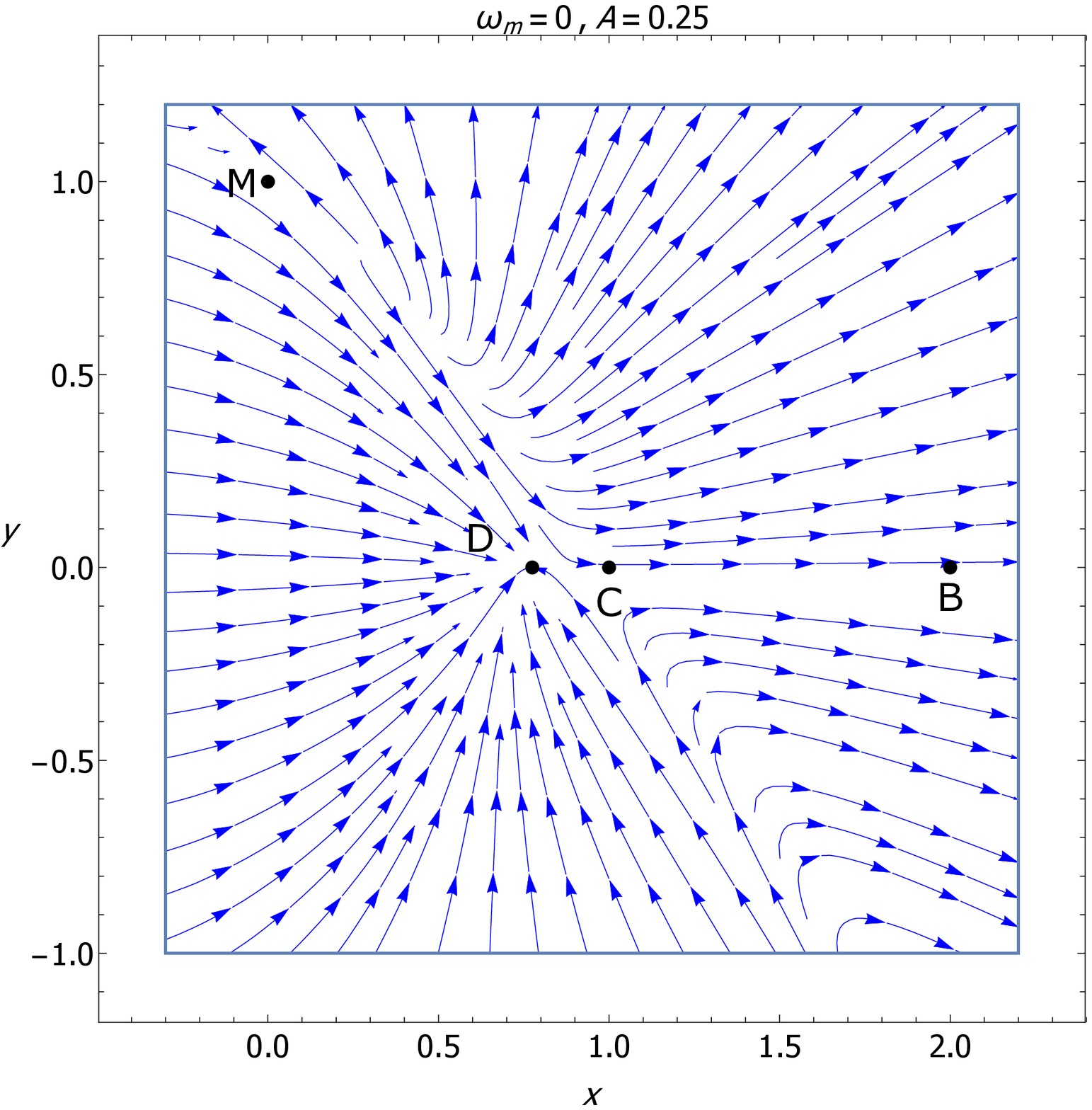}\label{fig:Stable_XY}}
	\caption{The figure shows the phase space projection of the autonomous system (\ref{autonomous}) on the H-x and x-y plane for the parameter values $\omega_{m}=0$, and $A=0.25$. In panel (a) in H-x plane, the set of critical points D is shown in the green colored curve which is attractor in the phase plane. The orange colored curves represent the set of points E is a saddle like (unstable) solution. Critical points M, B and C exhibit unstable solution. The panel (b) in x-y plane, the points M, B and C show the unstable (saddle like ) nature whereas D is stable there.}
	\label{phasespace-figure-HXY}
\end{figure}
\begin{figure}
	\centering
	\subfigure[]{%
		\includegraphics[width=8.5cm,height=8.5cm]{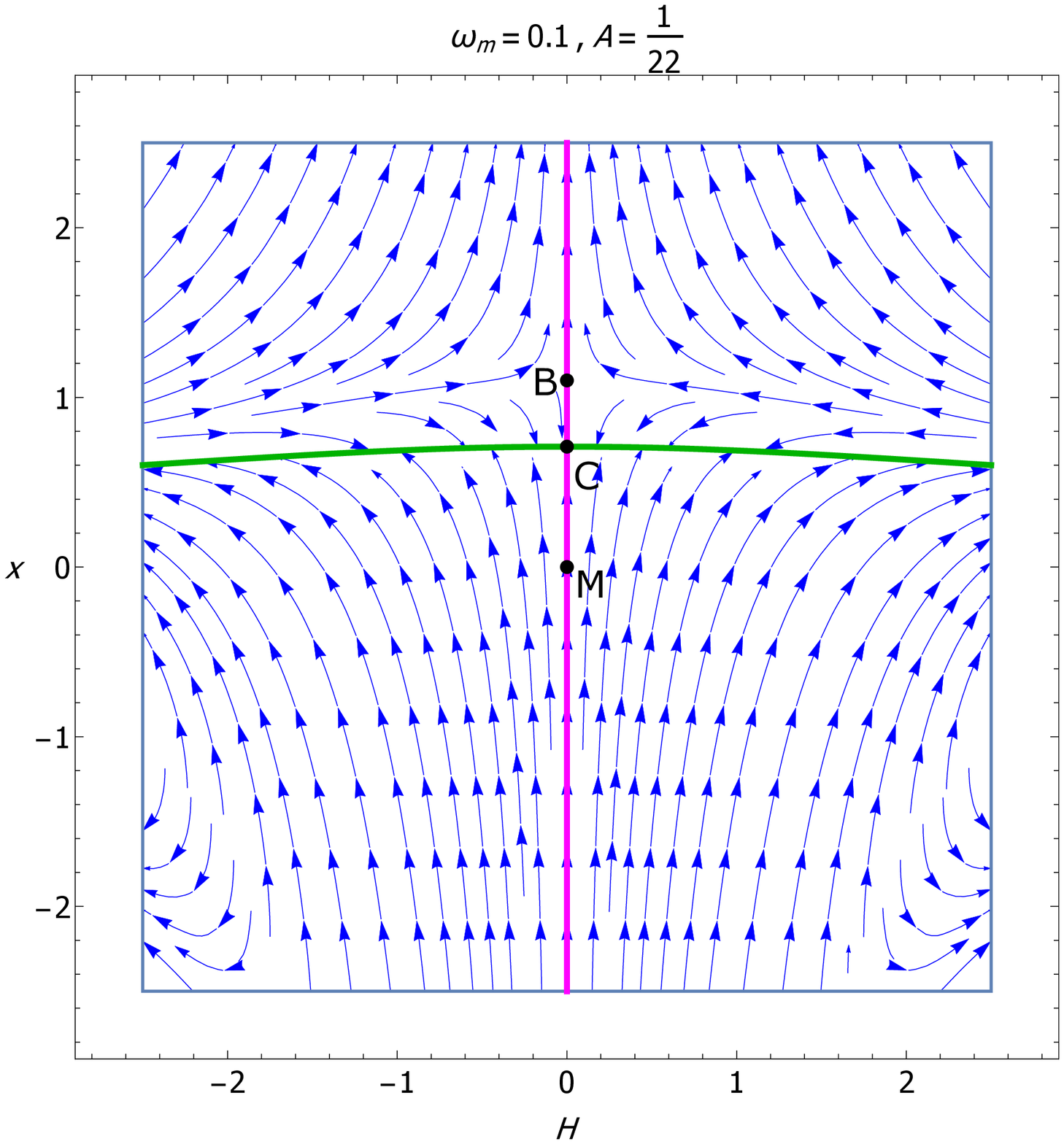}\label{fig:Stable_HX1}}
	\qquad
	\subfigure[]{%
		\includegraphics[width=8.5cm,height=8.5cm]{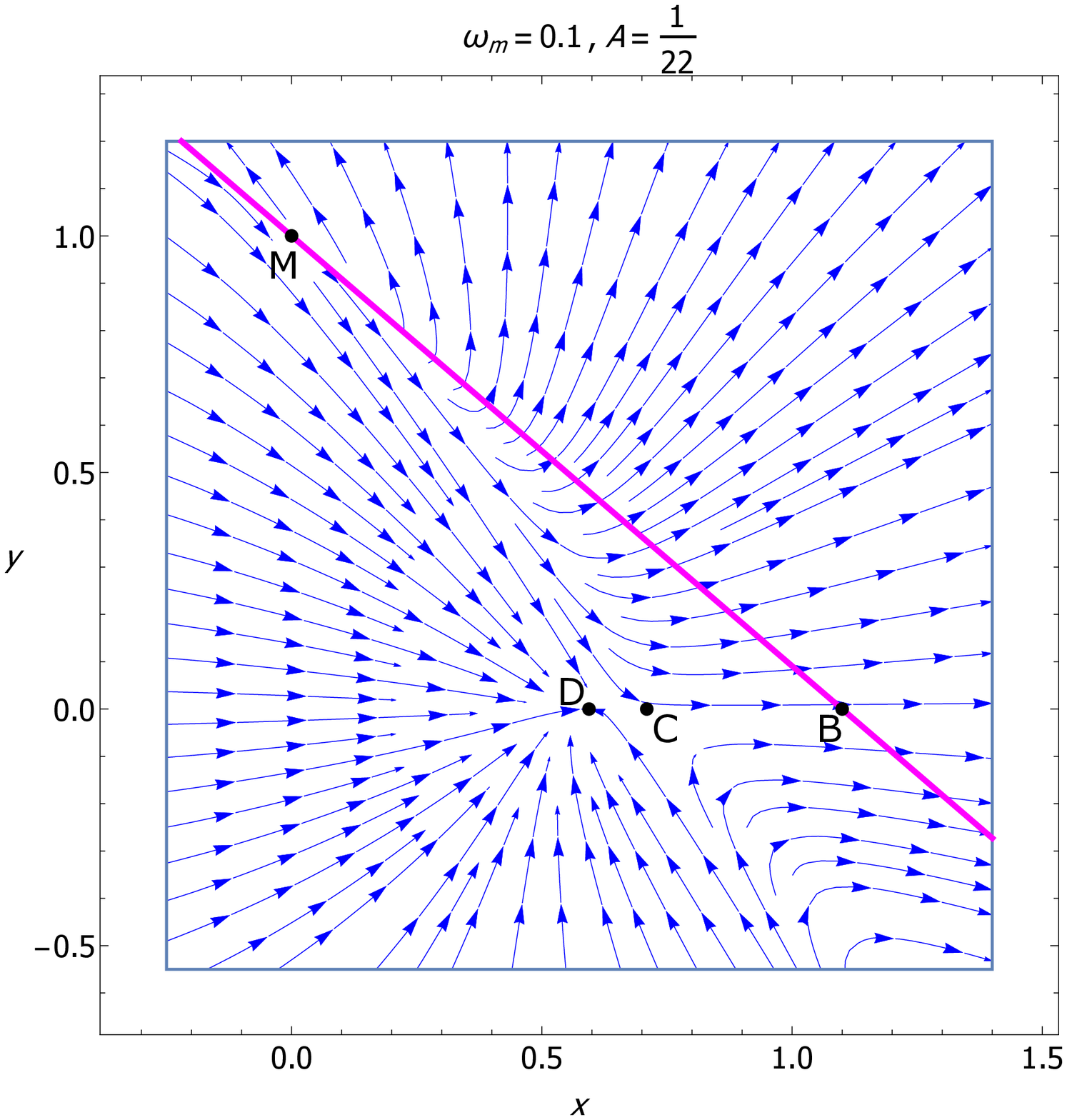}\label{fig:Stable_XY1}}
	\caption{The figure shows the projection of phase portrait on H-x and and x-y plane with the parameter values $\omega_{m}=0.1$, and $A=\frac{1}{22}$. In panel (a) and (b) the H-x phase plane and x-y phase plane exhibit that the critical points $M$, $B$ and $C$ are unstable saddle like solutions. Whereas the set of critical points $D$ with the green colored curve is an attractor in the phase plane and the set of critical points $F$ with magenta colored curve indicates the unstable (saddle-like) solution.}
	\label{phasespace-figure-HXY1}
\end{figure}


\begin{figure}
	\centering
	\subfigure[]{%
		\includegraphics[width=10cm,height=7cm]{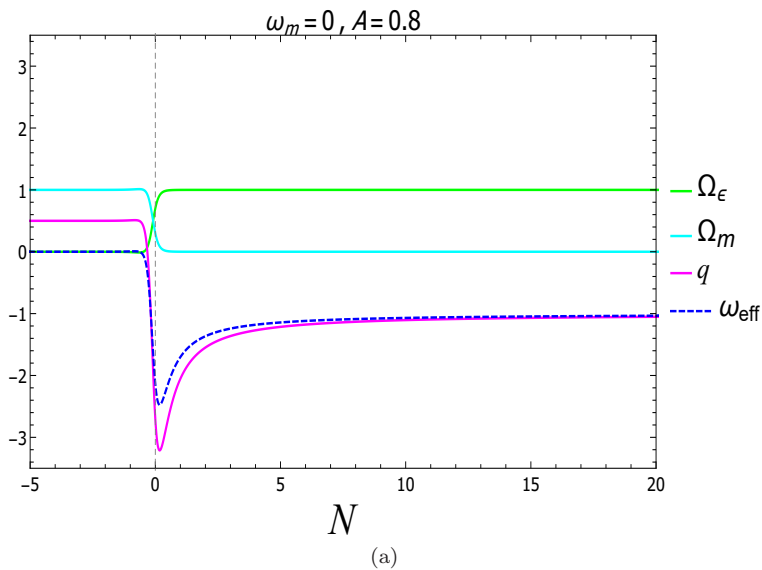}\label{fig:evolution}}
	\caption{Figure shows that the evolution of cosmological parameters like the DE density parameter for logotropic fluid $\Omega_{\epsilon}$, the density parameter for DM  $\Omega_{m}$, the effective equation of state parameter $\omega_{eff}$, and the decelerating parameter $q$ with respect to $N=\ln a$ of the autonomous system (\ref{autonomous}) with the parameter values $\omega_{m}=0$ and $A=0.8$. Numerical investigation shows that the ultimate fate of the universe is late time accelerated evolution in cosmological constant era followed by an intermediate matter dominated phase. Phantom crossing behavior is also reported here. }
	\label{phasespace-parameters-evolution}
\end{figure}


\subsection{Cosmology of the critical points}

In this section, we shall discuss the cosmological implications of the critical points obtained from the autonomous system (\ref{autonomous}) for the logotropic dark energy model. To perform dynamical analysis we introduce a dynamical variable $x$ in terms of $\rho$ normalized over Hubble scale. Here, the variable $x$ in phase space has the similar behavior as $\rho$ has in cosmological evolution, because there is one to one correspondence between the phase space variables and the cosmological variables. In particular, the cosmological parameters which already exist in terms of $\rho$ can be determined in terms of $x$ in dynamical analysis. Also, we have taken Hubble parameter as a dynamical variable and the phase space becomes 3D ($x-y-H$) system.\\

From the analysis of critical points, DM (perfect fluid model i.e., it can be dust, radiation or any other perfect fluid) dominated era of the universe can be achieved by the critical point $M$, where the universe is always decelerating and being a saddle like solution, it represents the intermediate phase of the evolution of the universe.\\ 

The universe near the critical point $B$ is completely dominated by logotropic fluid energy density ($\Omega_{\epsilon}=1$) of which is contributed by rest mass energy density. On restriction of parameter $A$, the logotropic fluid associated with the critical point $B$ can behave as dust, dark energy or any other exotic type matter. For example: $A \approx 0$ implies that the DE behaves as dust ($\omega_{\epsilon}\approx 0$) and the universe near the point evolves in a decelerated phase ($q\approx\frac{1}{2}, ~~\omega_{eff}\approx0$) having transient nature of evolution. On the other hand, for $A>\frac{1}{2}$, the logotropic fluid (DE) behaves as phantom like fluid ($\omega_{\epsilon}<-1$) and the accelerated universe near the critical point evolves in phantom regime ($\omega_{eff}<-1$), though the point is unstable there and the evolution is transient in nature. Figure (\ref{phasespace-figure-HXY}) for the parameter values $\omega_{m}=0$, $A=0.25$ exhibits that the decelerated saddle like solution $B$ representing the logotropic fluid (as stiff matter $\omega_{\epsilon}=1$) dominated decelerated ($\omega_{eff}=1$) universe. Also, the figure (\ref{phasespace-figure-HXY1}) for $\omega_{m}=0.1$ and $A=\frac{1}{22}$ shows that the critical point $B$ describes the decelerated logotropic fluid dominated univere where logotropic fluid mimicking as dark matter  ($\Omega_{\epsilon}=1$, $\omega_{\epsilon}=0.1$, $\omega_{eff}=0.1$). \\

Scaling solution represented by the set of points $F$ describes the decelerated transient phase of the universe. Logotropic fluid mimics here as the dark matter candidate in the form of perfect fluid (since $\omega_{\epsilon}=\omega_{m}$ where $0\leq \omega_{m}\leq 1$). It is dust for $\omega_{m}=0$ and radiation for $\omega_{m}=\frac{1}{3}$. The universe near the set evolves in dust and radiation era according as $\omega_{m}=0$ and $\omega_{m}=\frac{1}{3}$ (since $\omega_{eff}=\omega_{m}$).
The set of critical points F exhibits the feature of DM dominated decelerating phase of the universe for $x_c\longrightarrow 0$. In this case, the  point $M$ will become a particular case of $F$. On the other hand, for $x_c=1$ and $\omega_{m}=0$, the set $F$ describes the dust dominated decelerated universe where logotropic fluid acting as dust ($\Omega_{\epsilon}=1$, $\omega_{\epsilon}=0$, $\omega_{eff}=0$). Figure (\ref{phasespace-figure-HXY1}) shows that the set of points $F$ with magenta colored line evolves in unstable phase in the phase plane where decelerated universe is observed and logotropic fluid behaves as a perfect fluid model of dark matter ($\omega_{eff}=0.1$, $\omega_{\epsilon}=0.1$). Note that the critical points $M$, $B$ and $F$ in the figure do not represent the late time solutions there. As a result, there will be a phase transition from these points since after entering some trajectories into the points are immediately coming out of those points. In cosmological view point, these type of scenarios can be observed as intermediate phase of the universe.\\

The critical point $C$ represents completely logotropic fluid dominated ($\Omega_{\epsilon}=1$) universe evolving in cosmological constant era ($\omega_{eff}=-1$) where logotropic fluid behaves as cosmological constant like fluid ($\omega_{\epsilon}=-1$). From the analysis of dynamical systems, the point corresponds to a non-hyperbolic type critical point. Although, it has a 2D stable submanifold in the phase space for $A<\frac{1}{2}$, by performing numerical perturbation (see fig. (\ref{C-perturbation})), the point is saddle  (unstable) solution in the phase space.\\

After careful inspection, we observe\footnote{The authors are extremely grateful to the anonymous reviewer for drawing their attention to this important observation.} that this point is a de Sitter ($\Omega_{m}=0$, $\omega_{eff}=q=-1$) fixed point so that it may either describe dark energy in the form of a cosmological constant or even an inflationary point. Thus, the instability in the phase space may actually reveal the fact that the Universe is attracted to this fixed point, but eventually, due to its inherent instability, the Universe does not stay in this fixed point for too long. This can be important both mathematically as well as physically. Mathematically, it probably implies the existence of unstable manifolds in the phase space (see for example, in Refs. \cite{Chatzarakis2019,V.K. Oikonomou2019a,V.K. Oikonomou2020}), while physically, it can mean that the Universe ends its acceleration eventually. For an inflationary de Sitter point, this could mean graceful exit in simple words.\\

Late phase of the universe is characterized only by the set of critical points $D$. This is a normally hyperbolic set and conditionally stable in phase space. Here, the set is completely logotropic fluid dominated solution corresponding to an aacelerating universe and the acceleration is driven by cosmological constant. Logotropic fluid behaves as cosmological constant there. The figures (\ref{phasespace-figure-HXY} and \ref{phasespace-figure-HXY1}) show the stable attractor with green colored curve. Therefore, an accelerated evolution of the universe in cosmological constant era is achieved at late times by the set of points $D$ where the matter content is only the cosmological constant (logotropic fluid)   ($\omega_{\epsilon}=-1, \omega_{eff}=-1, q=-1$). \\

The set of critical points $E$ corresponds to an accelerated expansion of the universe, but it does not show the solution at late times. So, it is not physically interesting solution.\\

From the above discussions, several cosmological interesting scenarios are realized qualitatively for logotropic dark fluid model. We have obtained a unified description of evolution of the universe from matter dominated intermediate phase to dark energy dominated late phase. In the same parameter region, there exist a sequence of points such that: critical point $B$ and set $F$ (DM dominated solutions) $\longrightarrow$ set of points $D$ (DE dominated). It is worthy to note that logotropic fluid mimics as DM (dark matter) as well as DE (dark energy) (see figs. (\ref{phasespace-figure-HXY}) and (\ref{phasespace-figure-HXY1})).
In the late time scenario logotropic fluid behaves as the cosmological constant and at intermediate DM dominated decelerated phase of the universe, logotropic fluid behaves as DM in the form of perfect fluid  in the same parameter region. Also, a phantom crossing behavior can be achieved by the logotropic fluid in this model. These are shown numerically for this system in the
figure (\ref{phasespace-parameters-evolution}) where evolution of the cosmological parameters depict that the late phase of the universe is in cosmological constant era followed by an intermediate DM dominated era, and there also logotropic fluid crosses the phantom barrier.

\section{Summary and concluding remarks}
\label{sec4}

We have investigated the logotropic fulid taken as a DE candidate in the background of spatially flat FLRW metric in the perspective of dynamical system analysis. 
After some relevant approximations, logotropic fluid energy and pressure density have been modified with a power-law form of rest mass energy density. For the sake of dynamical analysis and to obtain a complete evolutionary picture of the system, we have considered a matter component in the form of DM. As a result, the modified Friedmann equation and the acceleration equation get modified and the respective autonomous system converted to a 3D system, including $H$ as a dynamical variable.\\ 

We have obtained several critical points from autonomous system. Linear stability theory have been employed to study the stability of the hyperbolic points. A non-hyperbolic point has been analyzed numerically to show the stability of that point. From the analysis of the critical points we have obtained results which are interesting from cosmological point of view.\\

The critical point $M$ corresponds to a matter dominated decelerated expansion of the universe. This point is saddle in the phase space. Therefore, it can successfully predict the intermediate phase of the universe.\\

The point $B$ has some attractive features such as it describes the universe dominated by logotropic fluid only which behaves as dust as well as phantom fluid depending on the parameter value $A$. In both the cases, the point is unstable (saddle) solution in the phase space. Although the universe enters into the phantom regime but still unstable there, showing its transient evolutionary scheme. The point is physically interesting only when it describes dust dominated decelerated phase of the universe driven by logotropic fluid mimicking dust fluid.\\

The non-hyperbolic critical point $C$ displays some physically interesting features. It represents a de Sitter ($\Omega_{m}=0$, $\omega_{eff}=q=-1$) fixed point. This implies that it may either describe dark energy in the form of a cosmological constant or even an inflationary point. So, the instability in the phase space may actually mean that the Universe is attracted to this fixed point. However, due to its inherent instability, the Universe does not stay in this fixed point for too long. This fact has important mathematical and physical implications. Mathematically, it probably implies the existence of unstable manifolds in the phase space and, physically, it can mean that the Universe ends its acceleration eventually. In simpler terms, this could mean graceful exit for an inflationary de Sitter point.\\

Scaling solution for this model is represented by the set of critical points $F$. It is always unstable (saddle like) in nature in the phase space. Logotropic fluid in this solution mimicking the dark matter behaves as dust or any other perfect fluid depending on $\omega_{m}$ (since $\omega_{\epsilon}=\omega_{m}$, where $0\leq\omega_{m}\leq1$). Logotropic fluid can never describes as DE candidate here. Therefore, the set of points $F$ is physically acceptable only when it predicts a matter dominated decelerated phase of the universe, where logotropic fluid acted as dark matter.\\

Finally, the set of points $D$ shows late time accelerated evolution of the universe successfully. This is a non-isolated set of points with exactly one vanishing eigenvalue indicating that it is a normally hyperbolic set. Imposing some constraints on parameters, it describes present accelerated expansion and the ultimate evolution for this case is in cosmological constant era. Here logotropic fluid behaves as a cosmological constant. See for example, in fig. (\ref{phasespace-parameters-evolution}) the evolution of the cosmological parameters depict the unified evolution of the universe from early dark matter dominated phase to late time dark energy dominated era and interestingly a phantom crossing behavior is also observed.\\

Therefore, we have obtained a DE dominated late time  accelerated evolution of the universe preceded by a matter dominated phase where logotropic fluid behaves as DE and DM at the same time and this is the interesting feature in logotropic model.

\section*{Acknowledgments}
The author Goutam Mandal is supported by UGC, Govt. of India through Junior Research Fellowship [Award Letter No. F.82-1/2018(SA-III)] for Ph.D. All the authors are indebted to the anonymous reviewer for his/her constructive comments which have helped to improve the quality of the manuscript significantly.


\end{document}